\documentclass{article}
\usepackage{latexsym}
\usepackage{amssymb}
\usepackage{amsmath,amsthm}
\usepackage[numbers,sort&compress]{natbib}
\usepackage{color}
\usepackage{amsfonts,amssymb}
\usepackage{mathrsfs}
\usepackage{dsfont}
\usepackage{graphicx}
\usepackage{subfigure}
\usepackage{float}
\usepackage{ulem}
\usepackage{color}
\usepackage{bm}

\newcommand{\sech}{\mbox{sech}}

\newtheorem{rem}{Remark}
\theoremstyle{definition}
\newtheorem{theo}{Theorem}[section]
\begin{document}
	\title{\bf Soliton dynamics to the Higgs equation and its multi-component generalization}
	\author{Wang Tang   \\
		 School of Mathematical Sciences, Shanghai Jiao Tong University, \\
		Shanghai 200240, P.R.\ China}
	\date{}
	\maketitle

\begin{abstract}
	In this paper, we study the coupled Higgs equation and its multi-component generalization based on the Hirota's direct method. One and two-soliton solutions of the coupled Higgs equation are derived by the perturbation approach. We express the $N$-soliton solutions in the form of Pfaffians and demonstrate that the $N$-component coupled Higgs equation turns out to be the Pfaffian identity. One and two-soliton solutions of the multi-component coupled Higgs equation are obtained from the Pfaffians. Starting from the explicit solutions, parallel solitons, periodic and nearly periodic interactions, elastic and inelastic collisions are investigated.
\end{abstract}


\section{Introduction}

Nonlinear integrable systems are of great significance in many fields of science, including Bose-Einstein condensate \cite{BE1,BE2}, nonlocal media \cite{media}, superfluids \cite{superfluid}, plasma \cite{plasma1,plasma2} and  optical systems \cite{optical1,optical2}. Different methods are developed to construct explicit solutions to nonlinear integrable systems, such as Darboux transformation method \cite{DT,DT1}, Hirota bilinear method \cite{BM,BM1}, inverse scattering method \cite{IS} and Riemann-Hilbert method \cite{RH}.

The following coupled Higgs equation is firstly investigated in \cite{Tajiri1}
\begin{align}\label{Eq-Higg2}
	\left\{
	\begin{aligned}
		&u_{tt}-u_{xx}-c u +r|u|^{2}u-2uv=0,  \\
		&v_{tt}+v_{xx}-r(|u|^{2})_{xx}=0,
	\end{aligned}
	\right.
\end{align}
where the function $v$ represents a real scalar meson field while $u$ is a complex scalar nucleon field. When $c<0$ and $r<0$, the system is the coupled nonlinear Klein-Gordon equation and it turns to be the coupled Higgs equation for $c>0$ and $r>0$. The system (\ref{Eq-Higg2}) describes a system of neutralscalar mesons interacting with conserved scalar nucleons in particle physics. This equation can be reduced from the coupled nonlinear Schr\"odinger equation \cite{Tajiri 1}. In this work, we focus on the coupled Higgs equation when $c>0$ and $r>0$.

The coupled system (\ref{Eq-Higg2}) is related to the classical Klein-Gordon equation $(c<0, r<0)$ or the Higgs equation $(c>0, r>0)$ \cite{V}
\begin{align}\label{Eq-Higg-single}
	&u_{tt}-u_{xx}-c u +r|u|^{2}u=0.
\end{align}
The equation (\ref{Eq-Higg-single}) has many important applications in a variety of fields, such as field theory, particle physics and electromagnetic waves. The equation  (\ref{Eq-Higg-single}) is related to the classical $u^4$ field theory in the physics of fields and elementary particles. It is shown that the coupled equation (\ref{Eq-Higg2}) has $N$-soliton solutions while the equation (\ref{Eq-Higg-single}) has only one-soliton solution. This fact is associated with that  equation (\ref{Eq-Higg2}) can be reduced to the second Painlev\'e equation while equation (\ref{Eq-Higg-single}) is not reducible \cite{Tajiri2}. In this sense, the coupled Higgs equation (\ref{Eq-Higg2}) has more abundant integrable properties than the system  (\ref{Eq-Higg-single}). Equation (\ref{Eq-Higg2}) is also related to the famous Davey-Stewartson (DS) equation
\begin{align}\label{DS}
	\left\{
	\begin{aligned}
		&\mathrm{i} \Psi_{\tilde{t}}-\Psi_{\tilde{x} \tilde{x}}+\Psi_{\tilde{y}\tilde{y}}+r|\Psi|^2 \Psi-2 \Psi \Phi=0, \\
		&\Phi_{\tilde{x} \tilde{x}}+\Phi_{\tilde{y} \tilde{y}}-r\left(|\Psi|^2\right)_{\tilde{x} \tilde{x}}=0 ,
	\end{aligned}
	\right.
\end{align}
via the following similar reduction \cite{Tajiri3}
\begin{align}\label{similar}
	\left\{
	\begin{aligned}
		&x=\tilde{x} / \tilde{t}, \quad t=\tilde{y} / \tilde{t}, \\
		&\Psi=\frac{1}{\tilde{t}} \exp \left[\frac{-\mathrm{i}\left(\tilde{x}^2-\tilde{y}^2+4c\right)}{4 \tilde{t}}\right] u(x, t), \\
		&\Phi=\frac{1}{\tilde{t}^2} v(x, t).
	\end{aligned}
	\right.
\end{align}
The DS equation (\ref{DS}) is a natural generalization of the nonlinear Schr\"odinger equation in the (2+1)-dimension.  (\ref{DS}) has the similar physical universality as the KP equation. This system is introduced by Davey and Stewartson to describe the evolution patterns of a wave-packet on water with finite depth in fluid dynamics \cite{DS}. The DS equation (\ref{DS}) is applicated in nonlinear optics, water waves and plasma physics and widely concerned nowadays \cite{DS1,DS2,DS3}. Considering the relationship between the coupled Higgs equation (\ref{Eq-Higg2}) and these two systems (\ref{Eq-Higg-single}) and (\ref{DS}) which are of great significance in physics, it is worthy for us to study the system (\ref{Eq-Higg2}).

For the coupled Higgs equation ($\ref{Eq-Higg2}$), integrability and $N$-soliton solutions are considered via Hirota's bilinear method \cite{Tajiri1} and homoclinic orbit method \cite{Hu}. Explicit solutions are found in the coupled Higgs system via exp-function method \cite{exp-function}. Travelling wave solutions in term of special functions are obtainted by using generalized algebraic method \cite{generalizedalgebraicmethod}.  Jacobi periodic wave solutions are considered with an algebraic method \cite{exactsolutions}. By using $G'/G$-expansion method,  trigonometric, rational function solutions, kinks, and periodic waves solutions are obtained \cite{G1,G2}. Doubly periodic solutions are investigated \cite{standingwave1,standingwave2}. Bifurcation of the travelling waves \cite{Bifurcationsoftravelingwave} and B\"acklund transformations \cite{Backlundtransformation} are also studied in the coupled Higgs equation.

Many integrable systems have their multi-component generalizations and explicit solutions with Pfaffian structures, such as the BKP hierarchy \cite{1,2}, the coupled mKdV equation \cite{3,4}, the discrete NLS equation \cite{5}, the integrable coupled dispersion equation \cite{6} and a (2+1)-dimensional differential-difference system \cite{7}. Motivated by these works, we consider a multi-component generalization of the coupled Higgs equation ($\ref{Eq-Higg2}$) in this work. The $N$-soliton solution is written in the compact form via Pfaffian technique. Though $N$-soliton solution is investigated \cite{Tajiri1}, the multi-component generalization and Pfaffian structure in the explicit $N$-soliton solution are not reported in the previous work. Moreover, we study further dynamic behaviours in both the single and multi-component coupled Higgs equation. Parallel solitons, elastic collisions, inelastic collisions, periodic interactions and nearly periodic interactions are found. Considering the special structure of the bilinear form of the coupled Higgs equation, we also draw a two soliton solution travelling in the opposite directions.

The paper is organized as follows. In section 2, we obtain one and two-soliton solutions of the coupled Higgs equation via Hirota's bilinear method and analyse the dynamic behaviours. $N$-soliton solutions of the multi-component coupled Higgs equation in the form of Pfaffian are given in section 3. In section 4, we study the soliton dynamics in the multi-component coupled Higgs equation. We make some conclusions in section 5.
\section{Soliton solutions of the coupled Higgs equation }
In this section, we find out the one soliton and two soliton solutions to the coupled Higgs equation (\ref{Eq-Higg2}) and analyse the dynamic behaviours of the solutions.
Upon the dependent variable transformation
\begin{equation}
	u=\frac{g}{f},\hspace{1em}v=2(\ln{f})_{xx},
\end{equation}
the nonlinear coupled Higgs equation \eqref{Eq-Higg2} transforms into the bilinear form
\begin{align}
	\left\{
	\begin{aligned}
		&(D^{2}_{t}-D^{2}_{x}-c)g \cdot f=0,  \\
		&(D^{2}_{t}+D^{2}_{x})f \cdot f=r|g|^2,
	\end{aligned}
	\right.
\end{align}
where $c,r$ are two arbitrary negative parameters.
The Hirota operator $D$ is defined as \cite{BM}
\begin{equation}
	D^{m}_{x}D^{k}_{t}a \cdot b=\Big(\frac{\partial}{\partial x}-\frac{\partial}{\partial x'}\Big)^{m}\Big(\frac{\partial}{\partial t}-\frac{\partial}{\partial t'}\Big)^{k}a(x,t)b(x',t')|_{x'=x,t'=t}.
\end{equation}

\subsection{One-soliton solution}

By expanding $f$ and $g$ in series of the parameter $\varepsilon$
\begin{align}
	f=1+\sum \limits_{j=1}^{\infty}\varepsilon^{2j}f^{(2j)},\hspace{1em}g=\sum \limits_{j=1}^{\infty}\varepsilon^{2j-1}g^{(2j-1)},
\end{align}
we obtain one-soliton solution
\begin{align}\label{Eq-1-soli-fg}
	g=\alpha_{1}e^{\eta_{1}},\hspace{1em}f=1+\frac{r|\alpha_{1}|^{2}}{2\Big({(p_{1}+\bar{p}_{1})}^{2}+{(q_{1}+\bar{q}_{1})}^{2}\Big)}e^{\eta_{1}+\bar{\eta}_{1}},
\end{align}
with $\eta_{1}=p_{1}x+q_{1}t$ and the dispersion relation $q_{1}^{2}-p_{1}^{2}-c=0$.
The solution $u$ and $v$ admit the representation
\begin{align}\label{Eq-1-soli-u}
	u=\frac{\alpha_{1}}{|\alpha_{1}|}\frac{\sqrt{2{(p_{1}+\bar{p}_{1})}^{2}+2{(q_{1}+\bar{q}_{1})}^{2}}}{2\sqrt{r}}e^{\frac{1}{2}(\eta_{1}-\bar{\eta}_{1})}\sech\Big(\frac{1}{2}(\eta_{1}+\bar{\eta}_{1})+\theta_{1}\Big),
\end{align}
and
\begin{align}\label{Eq-1-soli-v}
	v=2p_{1R}^{2}\sech^{2}\Big(\frac{1}{2}(\eta_{1}+\bar{\eta}_{1})+\theta_{1}\Big),
\end{align}
where $e^{2\theta_{1}}=\frac{r|\alpha_{1}|^{2}}{2\Big({(p_{1}+\bar{p}_{1})}^{2}+{(q_{1}+\bar{q}_{1})}^{2}\Big)}$. The solution is a bright soliton solution whose amplitude is $\frac{\sqrt{2{(p_{1}+\bar{p}_{1})}^{2}+2{(q_{1}+\bar{q}_{1})}^{2}}}{2\sqrt{r}}$, and the travelling velocity is $-\frac{q_{1R}}{p_{1R}}$, where $q_{1R}$ and $p_{1R}$ stand for the real parts of $q_{1}$ and $p_{1}$, respectively.
\begin{rem}
	We can see from (\ref{Eq-1-soli-fg}) that $u=\frac{g}{f}$ is nonsingular when $r>0$. However, there is a singularity in $u$ when $r<0$. It means that the coupled Higgs equation (\ref{Eq-Higg2}) with $c>0,r>0$ does not share every solutions with the coupled Klein-Gordon equation with $c<0,r<0$. In this work, we only focus on the coupled Higgs equation.
\end{rem}

\subsection{Two-soliton solution}

We obtain the 2-soliton solution to the coupled Higgs equations by standard Hirota's method, which takes the form of
\begin{align}\label{Eq-2 soliton}
	u=\frac{g_{1}+g_{3}}{1+f_{2}+f_{4}},\hspace{1em}v=2\Big(\ln{(1+f_{2}+f_{4})}\Big)_{xx},
\end{align}
where
\begin{align}
	&g_{1}=a_{1}e^{\eta_{1}}+a_{2}e^{\eta_{2}},\\
	&g_{3}=a_{12\bar1}|a_{1}|^{2}a_{2}e^{\eta_{1}+\eta_{2}+\bar{\eta}_{1}}+a_{12\bar2}|a_{2}|^{2}a_{1}e^{\eta_{1}+\eta_{2}+\bar{\eta}_{2}},\\
	&f_{2}= |a_{1}|^{2}a_{1\bar1}e^{\eta_{1}+\bar{\eta}_{1}}+|a_{2}|^{2}a_{2\bar2}e^{\eta_{2}+\bar{\eta}_{2}}+a_{1}\bar a_{2} a_{1\bar2}e^{\eta_{1}+\bar{\eta}_{2}}+a_{2}\bar a_{1} a_{2\bar1}e^{\eta_{2}+\bar{\eta}_{1}},\\
	&f_{4}= |a_{1}|^{2}|a_{2}|^{2}a_{12\bar1\bar2}e^{\eta_{1}+\eta_{2}+\bar{\eta}_{1}+\bar{\eta}_{2}},
\end{align}
with notations
\begin{figure}
	\begin{center}
		\begin{tabular}{cccc}
			\includegraphics[height=0.280\textwidth,angle=0]{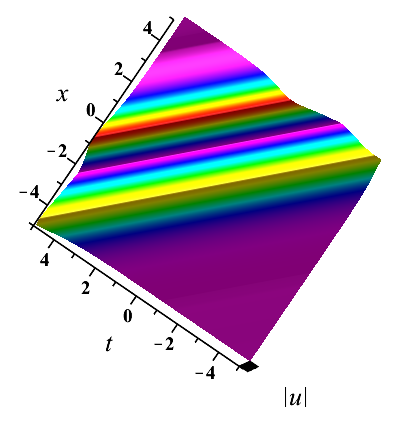} &
			\includegraphics[height=0.280\textwidth,angle=0]{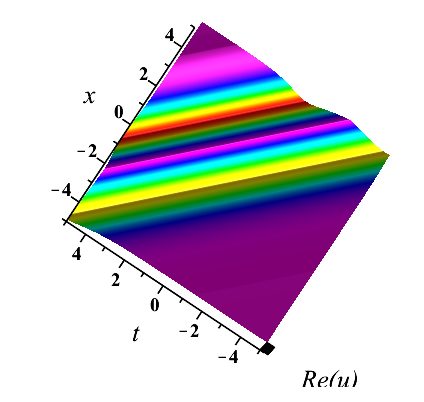} &\\
			\includegraphics[height=0.280\textwidth,angle=0]{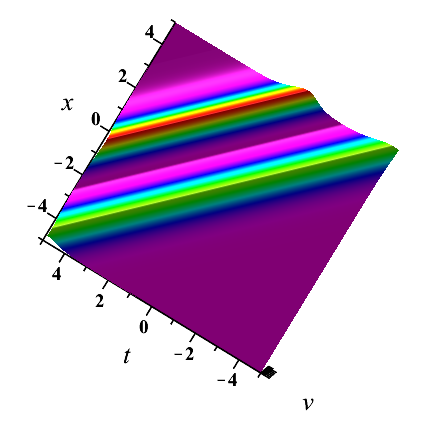} &
			\includegraphics[height=0.280\textwidth,angle=0]{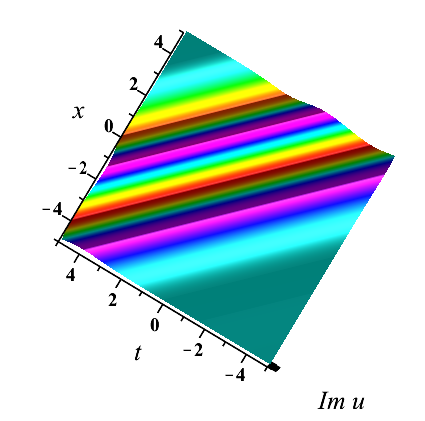} &
		\end{tabular}
	\end{center}
	\caption{Two parallel solitons in the coupled Higgs equation with $ p_{1}=1+\mathrm{i}, p_{2}=1+0.5\mathrm{i},c=0,r=2.$}
	\label{Fig-Two parallel solitons}
\end{figure}

\begin{figure}
	\begin{center}
		\begin{tabular}{cccc}
			\includegraphics[height=0.280\textwidth,angle=0]{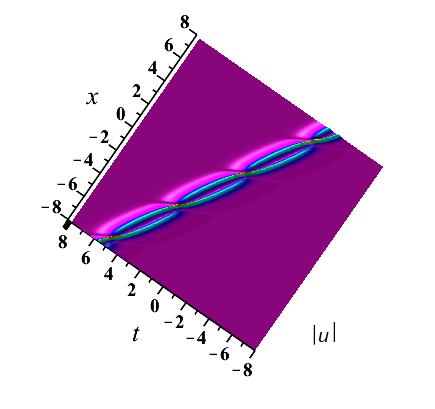} &
			\includegraphics[height=0.280\textwidth,angle=0]{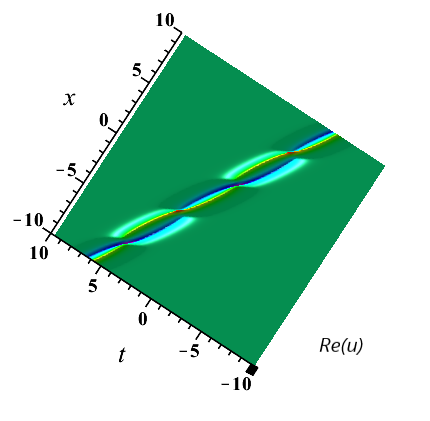} &\\
			\includegraphics[height=0.280\textwidth,angle=0]{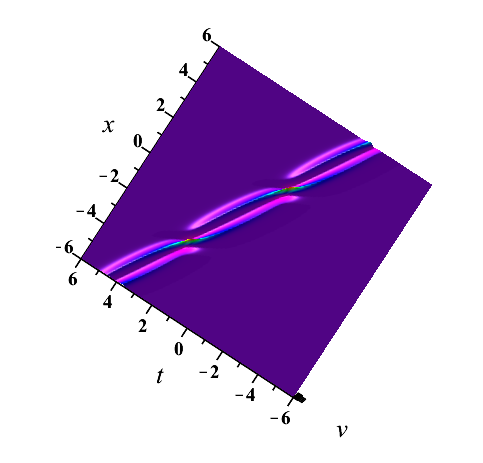} &
			\includegraphics[height=0.280\textwidth,angle=0]{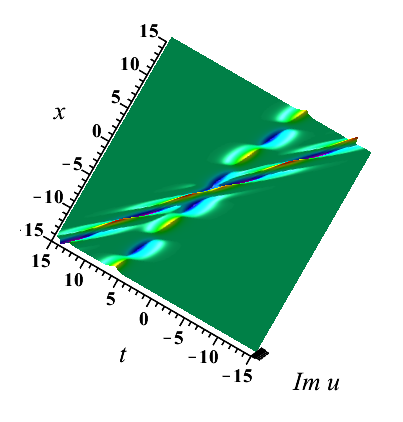}&
		\end{tabular}
	\end{center}
	\caption{Two-soliton periodic interaction in the coupled Higgs equation with $ p_{1}=3+\mathrm{i}, p_{2}=3-\mathrm{i},c=20,r=2.$}
	\label{Fig-Two-soliton periodic interaction}
\end{figure}

\begin{align}
	&a_{ij\bar k}=a_{ij}a_{i\bar k}a_{j\bar k},\hspace{1em}a_{ij\bar k\bar l}=a_{ij}a_{i\bar k}a_{j\bar k}a_{i\bar l}a_{j\bar l}a_{\bar k\bar l},\\
	&a_{ij}=\frac{2(q_{i}-q_{j})^{2}+2(p_{i}-p_{j})^{2}}{r},\\
	&a_{\bar i\bar j}=\frac{2(\bar q_{i}-\bar q_{j})^{2}+2(\bar p_{i}-\bar p_{j})^{2}}{r},\\
	&a_{i\bar j}=\frac{r}{2(q_{i}+\bar q_{j})^{2}+2(p_{i}+\bar p_{j})^{2}}.
\end{align}
and $\eta_{i}=p_{i}x+q_{i}t$, which satisfies the dispersion relation $q^2_{i}-p^2_{i}=c$.

Similar as the one-soliton situation, the amplitude and the velocity of each soliton are functions of the parameters $p_{i}$ and $q_{i}$.
So it leads to the collision of two solitons when $p_{i}$ and $q_{i}$ take different values.

We then check the asymptotic behaviour by representing the solutions in the form of sech function in order to prove the elastic collision without energy change. Then the phase shift can be easily recognized. We achieve the following asymptotic form for two-soliton solution (\ref{Eq-2 soliton})
$$ u\sim\left\{
\begin{aligned}
	&\frac{\sqrt{2{(p_{1}+\bar{p}_{1})}^{2}+2{(q_{1}+\bar{q}_{1})}^{2}}}{2\sqrt{r}}\sech\Big(\eta_{1R}+\frac{r_{12\bar{1}\bar{2}}-r_{2\bar{2}}}{2}\Big),\hspace{2em}\eta_{2R}\rightarrow +\infty,\eta_{1R}\sim O(1)\\
	&\frac{\sqrt{2{(p_{1}+\bar{p}_{1})}^{2}+2{(q_{1}+\bar{q}_{1})}^{2}}}{2\sqrt{r}}\sech\Big(\eta_{1R}+\frac{r_{1\bar{1}}}{2}\Big),\hspace{5em}\eta_{2R}\rightarrow -\infty,\eta_{1R}\sim O(1) \\
	&\frac{\sqrt{2{(p_{2}+\bar{p}_{2})}^{2}+2{(q_{2}+\bar{q}_{2})}^{2}}}{2\sqrt{r}}\sech\Big(\eta_{2R}+\frac{r_{12\bar{1}\bar{2}}-r_{2\bar{1}}}{2}\Big),\hspace{2em}\eta_{1R}\rightarrow +\infty,\eta_{2R}\sim O(1)\\
	&\frac{\sqrt{2{(p_{2}+\bar{p}_{2})}^{2}+2{(q_{2}+\bar{q}_{2})}^{2}}}{2\sqrt{r}}\sech\Big(\eta_{2R}+\frac{r_{2\bar{2}}}{2}\Big),\hspace{5em}\eta_{1R}\rightarrow -\infty,\eta_{2R}\sim O(1)
\end{aligned}
\right.
$$
and
$$ v\sim\left\{
\begin{aligned}
	&2p_{1R}^{2}\sech^{2}\Big(\eta_{1R}+\frac{r_{12\bar{1}\bar{2}}-r_{2\bar{2}}}{2}\Big),\hspace{1em}\eta_{2R}\rightarrow +\infty,\eta_{1R}\sim O(1)\\
	&2p_{1R}^{2}\sech^{2}\Big(\eta_{1R}+\frac{r_{1\bar{1}}}{2}\Big),\hspace{4em}\eta_{2R}\rightarrow -\infty,\eta_{1R}\sim O(1) \\
	&2p_{2R}^{2}\sech^{2}\Big(\eta_{2R}+\frac{r_{12\bar{1}\bar{2}}-r_{2\bar{1}}}{2}\Big),\hspace{1em}\eta_{1R}\rightarrow +\infty,\eta_{2R}\sim O(1)\\
	&2p_{2R}^{2}\sech^{2}\Big(\eta_{2R}+\frac{r_{2\bar{2}}}{2}\Big),\hspace{4em}\eta_{1R}\rightarrow -\infty,\eta_{2R}\sim O(1)
\end{aligned}
\right.
$$
with $e^{{r}_{2\bar 2}}=a_{2\bar 2},e^{{r}_{12\bar 1\bar 2}}=a_{12\bar 1\bar 2}$, where we take $a_{1}=a_{2}=1$.

We deduce that the amplitudes of $|u|$ and $v$ are $\frac{\sqrt{2{(p_{1}+\bar{p}_{1})}^{2}+2{(q_{1}+\bar{q}_{1})}^{2}}}{2\sqrt{r}},\frac{\sqrt{2{(p_{2}+\bar{p}_{2})}^{2}+2{(q_{2}+\bar{q}_{2})}^{2}}}{2\sqrt{r}}$ and $2p_{1R}^{2},2p_{2R}^{2}$ before and after the collisions respectively from the asymptotic analysis above. The amplitude, velocity and the shape of the waves keep the same before and after the collisions, which indicates that the collision is elastic without energy change. It permits a shift of the phase $\frac{r_{12\bar{1}\bar{2}}-r_{1\bar{1}}-r_{2\bar{2}}}{2}$ after each collision.

We can also make the observation that each wave travels in the speed of $-\frac{q_{iR}}{p_{iR}}$. When $a_{1}=a_{2}$ and $p_{i}$ are conjugate to each other, two waves might be close enough and oscillate periodically. The periodicity of the imaginary and real parts of both solitons cause the periodic phenomena.

By calculating $\frac{\eta_{i}-\eta^{*}_{i}}{2}$ we achieved the collision period. We denote the velocity of each soliton by $v_1=-\frac{q_{1R}}{p_{1R}}$ and  $v_2=-\frac{q_{2R}}{p_{2R}}$, respectively. Note that
\begin{align}
	\eta_{i}=p_{i}x+q_{i}t,\hspace{1em}q_{i}^{2}-p_{i}^{2}=c,\hspace{1em}\frac{\eta_{i}-\bar\eta_{i}}{2}=\mathrm{i}\operatorname{Im}(q_{i})t+\mathrm{i}\operatorname{Im}\Bigg(\sqrt{q_{i}^{2}-c}\Bigg)x,
\end{align}
and $x=v_i t+Const$. We set $Const=0$ without loss of generality. Then we have
\begin{align}
	\frac{\eta_{i}-\bar\eta_{i}}{2}=\mathrm{i}\left(\operatorname{Im}(q_{i})-\operatorname{Im}\Bigg(\sqrt{q_{i}^{2}-c}\Bigg)\frac{q_{iR}}{p_{iR}}\right)t.
\end{align}
Therefore, the collision period $T$ has the expression
\begin{align}\label{Eq-periodicity}
	T=\left|\frac{2\pi}{\operatorname{Im}(q_{1})-\operatorname{Im}(q_{2})-\operatorname{Im}\Bigg(\sqrt{q_{1}^{2}-c}\Bigg)\frac{q_{1R}}{p_{1R}}+\operatorname{Im}\Bigg(\sqrt{q_{2}^{2}-c}\Bigg)\frac{q_{2R}}{p_{2R}}}\right|.
\end{align}

\begin{figure}
	\begin{center}
		\begin{tabular}{cc}
			\includegraphics[height=0.330\textwidth,angle=0]{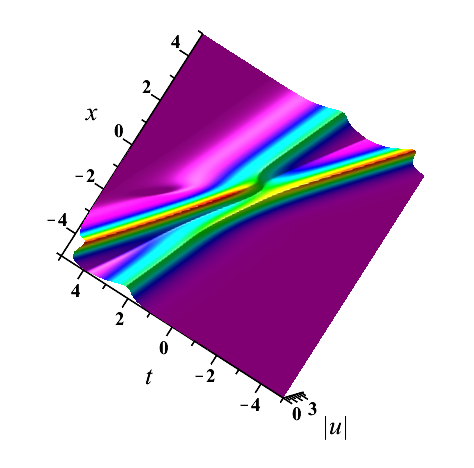} &
			\includegraphics[height=0.330\textwidth,angle=0]{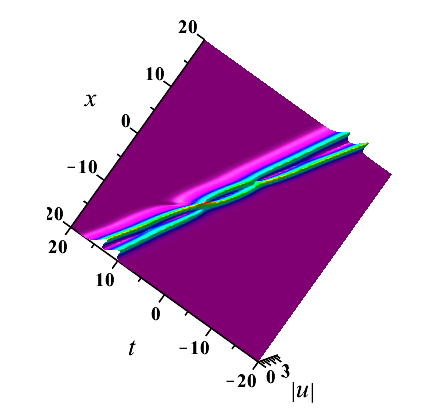} \\
			$(a) \quad  $ & $(b)\quad   $
		\end{tabular}
	\end{center}
	\caption{(a) Elastic collision when $ p_{1}=3+\mathrm{i}, p_{2}=3.4-\mathrm{i},c=20,r=5$. (b) Nearly periodic interaction when $ p_{1}=3+3\mathrm{i}, p_{2}=1+0.9\mathrm{i},c=6,r=3.$}
	\label{Fig-Elastic collision}
\end{figure}

\begin{figure}
	\begin{center}
		\begin{tabular}{ccc}
			\includegraphics[height=0.280\textwidth,angle=0]{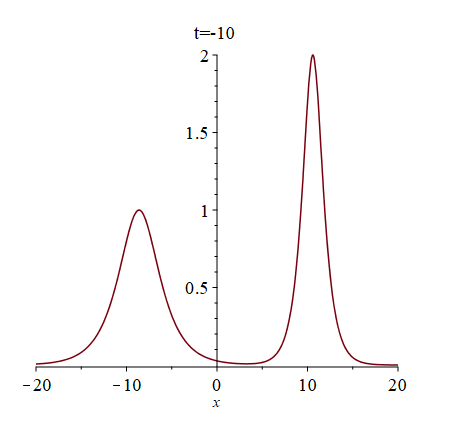} &
			\includegraphics[height=0.280\textwidth,angle=0]{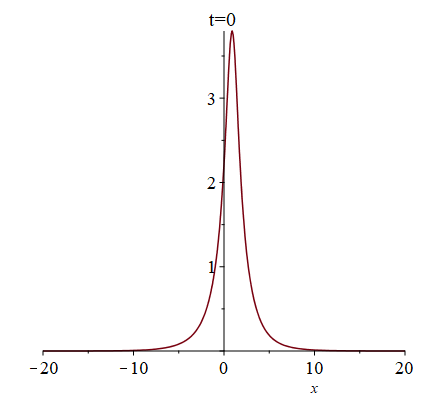} &
			\includegraphics[height=0.280\textwidth,angle=0]{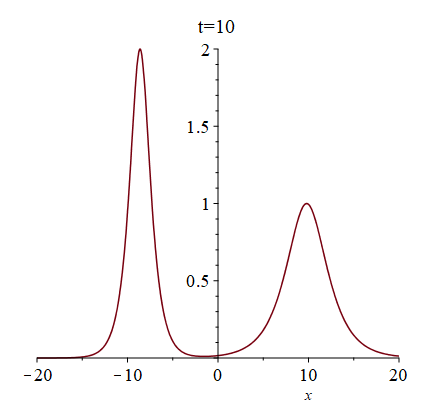} \\
		\end{tabular}
	\end{center}
	\caption{Two-soliton that travel in the opposite direction of the coupled Higgs equation with $ p_{1}=1+\mathrm{i}, p_{2}=\frac{1+\sqrt{7}\mathrm{i}}{2},c=0,r=1.$}
	\label{Fig-Opposite direction}
\end{figure}

Firstly, we consider the special case when $c=0$. Though $c=0$ is out of the parameter range of the coupled Higgs equation (\ref{Eq-Higg2}), we still consider it as a limit case. In this case, the two solitons travel in the same speed and thus the solution is consisted of two parallel solitons. In Fig. \ref{Fig-Two parallel solitons}, we depict the two parallel solitons when we set the parameters as $ p_{1}=1+\mathrm{i}, p_{2}=1+0.5\mathrm{i},c=0,r=2$. When $c\neq 0$, periodic interaction is shown in Fig. \ref{Fig-Two-soliton periodic interaction} when $ p_{1}=3+\mathrm{i}, p_{2}=3-\mathrm{i},c=20,r=2$ and the periodicity satisfies the period formula (\ref{Eq-periodicity}). We see the elastic interaction with a phase shift when $ p_{1}=3+\mathrm{i}, p_{2}=3.4-\mathrm{i},c=20,r=5$ in Fig. \ref{Fig-Elastic collision}(a) and
a nearly periodic interaction when $ p_{1}=3+3\mathrm{i}, p_{2}=1+0.9\mathrm{i},c=6,r=3$ in Fig. \ref{Fig-Elastic collision}(b). In the nearly periodic interaction, the two solitons oscillate together for a period of time and finally travel apart. The velocity and amplitude keep no change before and after the collision in both cases.

To draw another kind of 2-soliton solution, due to the special structure of the equation, the coefficient of $\varepsilon$ is not trivial if we set $\eta_{1}=p_{1}x+p_{1}t,\eta_{2}=p_{2}x-p_{2}t$. We find a two-soliton solution travelling in the opposite directions. This solution can be written in the following form

\begin{align}
	u=\frac{g_{1}+g_{3}}{1+f_{2}+f_{4}},\hspace{1em}v=2\Big(\ln{(1+f_{2}+f_{4})}\Big)_{xx},
\end{align}
where
\begin{align*}
	&g_{1}=a_{1}e^{\eta_{1}}+a_{2}e^{\eta_{2}},\\
	&f_{2}=\frac{a_{1}\bar{a}_{1}}{4(p_{1}+\bar{p}_1)^{2}}e^{\eta_{1}+\bar{\eta}_{1}}+\frac{a_{2}\bar{a}_{2}}{4(p_{2}+\bar{p}_{2})^{2}}e^{\eta_{2}+\bar{\eta}_{2}}
	+\frac{a_{1}\bar{a}_{2}}{4(p_{1}^{2}+{\bar{p}_{2}^{2}})}e^{\eta_{1}+\bar{\eta}_{2}}+\frac{a_{2}\bar{a}_{1}}{4(p_{2}^{2}+{\bar{p}_{1}^{2}})}e^{\eta_{2}+\bar{\eta}_{1}},\\
	&g_{3}=\frac{1}{4}|a_{1}|^{2}a_{2}\left(\frac{1}{p_{1}+\bar{p}_{1}}+\frac{p_{1}-\bar{p}_{1}}{p_{1}+\bar{p}_{1}}\frac{1}{p_{2}^{2}+{\bar{p}_{1}^{2}}}\right)e^{\eta_{1}+\eta_{2}+\bar\eta_{1}}\\
	&\quad\quad+\frac{1}{4}|a_{2}|^{2}a_{1}\left(\frac{1}{p_{2}+\bar{p}_{2}}+\frac{p_{2}-\bar{p}_{2}}{p_{2}+\bar{p}_{2}}\frac{1}{p_{1}^{2}+{\bar {p}_{2}^{2}}}\right)e^{\eta_{1}+\eta_{2}+\bar\eta_{2}},\\
	&f_{4}=\frac{|a_{1}|^{2}|a_{2}|^{2}}
	{16\Big((p_{1}+\bar{p}_{1})^{2}+(p_{2}+\bar{p}_{2})^{2}\Big)}A,\\
\end{align*}
and
\begin{align*}
	A=&\left(\frac{1}{(p_{1}+\bar{p}_{1})^2}+\frac{1}{(p_{2}+\bar{p}_{2})^2}+
	\Big(\frac{p_{2}-\bar{p}_{2}}{p_{2}+\bar{p}_{2}}-\frac{p_{1}-\bar{p}_{1}}{p_{1}+\bar{p}_{1}}\Big)\Big(\frac{1}{p_{2}^{2}+{\bar{p}_{1}^{2}}}-\frac{1}{p_{1}^{2}+{\bar{p}_{2}^{2}}}\Big)\right.\\
	&\left.-\frac {(p_{1}-\bar{p}_{1})^{2}+(p_{2}-\bar{p}_{2})^{2}}{(p_{1}^{2}+\bar p_{2}^{2})(p_{2}^{2}+\bar p_{1}^{2})}\right) e^{\eta_{1}+\eta_{2}+\bar\eta_{1}+\bar\eta_{2}},
\end{align*}
under this circumstance.
The two solitons travel in the opposite directions and a much higher amplitude is attained when collision. The amplitudes and shapes of the two solitons stay the same before and after the collision. There are no phase shifts under this circumstance.

\section{Pfaffian expression of the solutions to the multi-component coupled Higgs equation}
\subsection{One-soliton solutions to the multi-component coupled Higgs equation}
\quad \quad In this section, we give the multi-component generalization to the coupled Higgs equation and write the $N$-soliton solutions of the multi-component coupled Higgs equation in the form of Pfaffians. As is well-known, many famous equations have their multi-component forms. Thus it is natural for us to consider the multi-component form of the coupled Higgs equation. We consider the following multi-component generalization of the coupled Higgs equation (\ref{Eq-Higg2})
\begin{equation}
	\left\{
	\begin{aligned}
		&u_{i,tt}-u_{i,xx}-cu_{i}+r\Big(\sum_{j=1}^{n}\delta_{j}{|u_{j}|}^{2}\Big){u}_{i}-2{u}_{i}v=0,\\
		&v_{tt}+v_{xx}-r\Big(\sum_{j=1}^{n}\delta_{j}{|u_{j}|}^{2}\Big)_{xx}=0,
	\end{aligned}
	\right.
\end{equation}
where $\delta_{j}$ are arbitrary real numbers.
By taking the dependent variable transformation $u_{j}=\frac{g_{j}}{f}$ and $v=2{(\ln f)}_{xx}$ we derive the bilinear form of the multi-component system, where we assume that $g$ is a complex function while $f$ is a real function with $c,r$ two positive parameters,
\begin{equation}\label{multi-Higgs}
	\left\{
	\begin{aligned}
		&(D^{2}_{t}-D^{2}_{x}-c)g_{i}\cdot f=0,\\
		& (D^{2}_{t}+D^{2}_{x})f\cdot f=r\sum_{i=1}^{n}\delta_{i}{|g_{i}|}^{2}.
	\end{aligned}
	\right.
\end{equation}
We can express $N$-soliton solutions of the multi-component coupled Higgs equation in the Pfaffian form in the following theorem.
\begin{theo}
	The Pfaffians
	\begin{align}
		&f=(a_{1},a_{2},\dots,a_{N},a^*_{1},a^*_{2},\dots,a^*_{N},b^*_{N},\dots,b^*_{2},b^*_{1},b_{N},\dots,b_{2},b_{1} ),\\
		&g_{k}=(d_{0},\beta_{k},a_{1},a_{2},\dots,a_{N},a^*_{1},a^*_{2},\dots,a^*_{N},b^*_{N},\dots,b^*_{2},b^*_{1},b_{N},\dots,b_{2},b_{1} ),
	\end{align}
	with the elements of the Pfaffians defined by
	\begin{align*}
		&(a_{i},a_{j})=\frac{p_{i}-p_{j}}{q_{i}+q_{j}}e^{\eta_{i}+\eta_{j}},\hspace{1em}(a^*_{i},a^*_{j})=\frac{\bar{p}_{i}-\bar{p}_{j}}{\bar{q}_{i}+\bar{q}_{j}}e^{\bar{\eta}_{i}+\bar{\eta}_{j}},\hspace{1em}(a_{i},a^*_{j})=\frac{p_{i}-\bar{p}_{j}}{q_{i}+\bar{q}_{j}}e^{\eta_{i}+\bar{\eta}_{j}},\\
		&(b_{j},b_{i})=(b^*_{j},b^*_{i})=0,\hspace{1em}(b_{i},b^*_{j})=-\sum\limits_{k=1}^{N}\frac{r}{4}\frac{\delta_{k}\alpha_{i}^{(k)}\bar{\alpha}_{j}^{(k)}}{p_{i}q_{i}-\bar{p}_{j}\bar{q}_{j}},\hspace{1em}(a_{i},b_{j})=(a^*_{i},b^*_{j})=\delta_{ij},\\
		&(b^*_{j},\beta_{k})=(a_{i},\beta_{k})=(a^*_{j},\beta_{k})=0,\hspace{1em}(a_{i},b^*_{j})=(a^*_{j},b_{i})=0,\hspace{1em}(b_{j},\beta_{k})=\alpha_{j}^{(k)},\hspace{1em}(b^*_{j},\beta^*_{k})=\bar\alpha_{j}^{(k)},\\
		&(b_{j},\beta^*_{k})=(a_{i},\beta^*_{k})=(a^*_{j},\beta^*_{k})=0,\hspace{1em}(d_{0},a_{i})=e^{\eta_{i}},\hspace{1em}(d_{0},a^*_{i})=e^{\bar{\eta}_{i}},\hspace{1em}(d_{1},a_{i})=p_{i}e^{\eta_{i}},\\
		&(d_{1},a^*_{i})=\bar{p}_{i}e^{\bar{\eta}_{i}}\hspace{1em}(d_{2},a_{i})=p^{2}_{i}e^{\eta_{i}},\hspace{1em}(d_{2},a^*_{i})=\bar{p}^{2}_{i}e^{\bar{\eta}_{i}},\hspace{1em}(d_{-1},a_{i})=q_{i}e^{\eta_{i}},\hspace{1em}(d_{-1},a^*_{i})=\bar{q}_{i}e^{\bar{\eta}_{i}},\\
		&(d_{-2},a_{i})=q^{2}_{i}e^{\eta_{i}},\hspace{1em}(d_{-2},a^*_{i})=\bar{q}^{2}_{i}e^{\bar{\eta}_{i}},\hspace{1em}(\rho,a_{i})=p_{i}q_{i}e^{\eta_{i}},\hspace{1em}(\rho,a^*_{i})=\bar{p}_{i}\bar{q}_{i}e^{\bar{\eta}_{i}},\\
		&(d_{l},b_{i})=(d_{l},b^*_{i})=(d_{l},\beta_{k})=(d_{l},d_{j})=0,\hspace{1em}(\rho,b_{i})=(\rho,b^*_{i})=(\rho,\beta_{k})=(\rho,d_{j})=0,
	\end{align*}
	solves the bilinear multi-component coupled Higgs equation (\ref{multi-Higgs}), where $\eta_{i}=p_{i}x+\sqrt{c+p^{2}_{i}}t.$
\end{theo} 
For simplicity, we introduce the notations that
\begin{align}
	f=(\bullet),\hspace{1em}g_{k}=(d_{0},\beta_{k},\bullet).
\end{align}
Then the derivatives of $f$ and $g_{k}$ can be calculated as
\begin{align}
	&\frac{\partial}{\partial x}f=(d_{0},d_{-1},\bullet),\hspace{1em}\frac{\partial}{\partial t}f=(d_{0},d_{1},\bullet),\hspace{1em}\frac{\partial}{\partial x}g_{k}=(d_{1},\beta_{k},\bullet),\hspace{1em}\frac{\partial}{\partial t}g_{k}=(d_{-1},\beta_{k},\bullet),\\
	&\frac{\partial^{2}}{\partial x^{2}}f=(d_{0},\rho,\bullet)-(d_{-1},d_{1},\bullet),\hspace{1em}\frac{\partial^{2}}{\partial t^{2}}f=(d_{0},\rho,\bullet)+(d_{-1},d_{1},\bullet),\\
	&\frac{\partial^{2}}{\partial x^{2}}g_{k}=(d_{2},\beta_{k},\bullet)+(d_{0},d_{-1},d_{1},\beta_{k},\bullet),\hspace{1em}\frac{\partial^{2}}{\partial t^{2}}g_{k}=(d_{-2},\beta_{k},\bullet)-(d_{0},d_{-1},d_{1},\beta_{k},\bullet).
\end{align}

Thanks to the Pfaffian identity, we have that
\begin{equation}\label{Eq-first-identity}
	\begin{split}
		(D_{t}^{2}-D_{x}^{2}-c)g_{k}\cdot f
		&=(g_{k,tt}f-2g_{k,t}f_{t}+g_{k}f_{tt})-(g_{k,xx}f-2g_{k,x}f_{x}+g_{k}f_{xx})-cg_{k}f\\
		&=\Big((d_{-2},\beta_{k},\bullet)-(d_{2},\beta_{k},\bullet)-c(d_{0},\beta_{k},\bullet)\Big)(\bullet)\\
		&+2(d_{0},\beta_{k},\bullet)(d_{-1},d_{1},\bullet)+2(d_{1},\beta_{k},\bullet)(d_{0},d_{-1},\bullet)\\
		&\quad-2(d_{-1},\beta_{k},\bullet)(d_{0},d_{1},\bullet)-2(d_{0},d_{-1},d_{1},\beta_{k},\bullet)(\bullet)\\
		&=\Big((d_{-2},\beta_{k},\bullet)-(d_{2},\beta_{k},\bullet)-c(d_{0},\beta_{k},\bullet)\Big)(\bullet)\\
		&=\sum\limits_{i=1}^{N}(-1)^{i}\Big((d_{-2},a_{i})-(d_{2},a_{i})-c(d_{0},a_{i})\Big)(\beta_{k},\hat a_{i})(\bullet)\\
		&+\sum\limits_{i=1}^{N}(-1)^{i}\Big((d_{-2},a^*_{i})-(d_{2},a^*_{i})-c(d_{0},a^*_{i})\Big)(\beta_{k},\hat a^*_{i})(\bullet)\\
		&=\sum\limits_{i=1}^{N}(-1)^{i}(q_{i}^{2}-p_{i}^{2}-c)e^{\eta_{i}}(\beta_{k},\hat a_{i})(\bullet)+\sum\limits_{i=1}^{N}(-1)^{i}(\bar q_{i}^{2}-\bar p_{i}^{2}-c)e^{\bar\eta_{i}}(\beta_{k},\hat a^*_{i})(\bullet)\\
		&=0,
	\end{split}
\end{equation}
where the dispersion relation is utilized. (\ref{Eq-first-identity}) gives a proof to the first bilinear identity. For the second one, we calculate the right side of the second bilinear identity as
\begin{equation}
	\begin{split}
		r\sum \limits_{k=1}^{N}\delta_{k}g_{k}g_{k}^{*} &=r\sum \limits_{k=1}^{N}\delta_{k}(d_{0},\beta_{k},\bullet)(d_{0},\beta^*_{k},\bullet)\\
		&=r\sum \limits_{k=1}^{N}\delta_{k}\sum \limits_{i=1}^{N}(-1)^{i+1}(b_{i},\beta_{k})(d_{0},\hat{b}_{i})\sum \limits_{j=1}^{N}(-1)^{N+j+1}(b^*_{j},\beta^*_{k})(d_{0},\hat{b}^*_{j})\\
		&=\sum \limits_{i,j=1}^{N}(-1)^{i+j+N}(r\sum \limits_{k=1}^{N}\delta_{k}\alpha_{i}^{(k)}\bar{\alpha}_{j}^{(k)})(d_{0},\hat{b}_{i})(d_{0},\hat{b}^*_{j})\\
		&=-4\sum \limits_{i,j=1}^{N}(-1)^{i+j+N}(p_{i}q_{i}-\bar{p}_{j}\bar{q}_{j})(b_{i},b^*_{j})(d_{0},\hat{b}_{i})(d_{0},\hat{b}^*_{j})\\
		&\overset{\triangle}{=}I.
	\end{split}
\end{equation}
Considering that
\begin{equation}
	\begin{split}
		0 &=(d_{0},b^*_{j},a_{1},a_{2},\dots,a_{N},a^*_{1},a^*_{2},\dots,a^*_{N},b^*_{N},\dots,b^*_{2},b^*_{1},b_{N},\dots,b_{2},b_{1} )\\
		&=\sum \limits_{i=1}^{N}(-1)^{i+N}(b^*_{j},a^*_{i})(d_{0},\hat{a^*_{i}})+\sum \limits_{i=1}^{N}(-1)^{i-1}(b^*_{j},b_{i})(d_{0},\hat{b}_{i})\\
		&=(-1)^{j+N+1}(d_{0},\hat{a_{j}^*})+\sum \limits_{i=1}^{N}(-1)^{i-1}(b^*_{j},b_{i})(d_{0},\hat{b}_{i}), \\
	\end{split}
\end{equation}
we have the following expressions
\begin{equation}
	\begin{split}
		&\sum \limits_{i,j=1}^{N}-\bar{p}_{j}\bar{q}_{j}(-1)^{i+j+N}(d_{0},\hat{b}_{i})(d_{0},\hat{b}^*_{j})(b_{i},b^*_{j})\\
		&=\sum \limits_{j=1}^{N}-\bar{p}_{j}\bar{q}_{j}(-1)^{j+N}(d_{0},\hat{b}^*_{j})\sum \limits_{i=1}^{N}(-1)^{i}(d_{0},\hat{b}_{i})(b_{i},b^*_{j}) \\
		&=\sum \limits_{j=1}^{N}-\bar{p}_{j}\bar{q}_{j}(-1)^{j+N}(d_{0},\hat{b}^*_{j})[(-1)^{j+N}(d_{0},\hat{a}^*_{j})]\\
		&=-\sum \limits_{j=1}^{N}\bar{p}_{j}\bar{q}_{j}(d_{0},\hat{b}^*_{j})(d_{0},\hat{a}^*_{j}).
	\end{split}
\end{equation}
Simialrly, we can deduce that
\begin{align}
	\sum \limits_{i,j=1}^{N}p_{j}q_{j}(-1)^{i+j+N}(d_{0},\hat{b}_{i})(d_{0},\hat{b}^*_{j})(b_{i},b^*_{j})=-\sum \limits_{i=1}^{N}{p}_{i}{q}_{i}(d_{0},\hat{a}_{i})(d_{0},\hat{b}_{i}),
\end{align}
and further we come to the following identity
\begin{align}\label{Eq-I}
	I=4\sum \limits_{i=1}^{N}\Big({p}_{i}{q}_{i}(d_{0},\hat{a}_{i})(d_{0},\hat{b}_{i})+\bar{p}_{i}\bar{q}_{i}(d_{0},\hat{a}^*_{i})(d_{0},\hat{b}^*_{i})\Big).
\end{align}
For simplicity, we denote ${a}^*_{i},{b}^*_{i},\bar{p}_{i},\bar{q}_{i}$ as ${a}_{N+i},{b}_{N+i},p_{N+i},q_{N+i}$ respectively hereafter. So we can rewrite (\ref{Eq-I}) as
\begin{align}
	I=4\sum \limits_{i=1}^{2N}{p}_{i}{q}_{i}(d_{0},\hat{a}_{i})(d_{0},\hat{b}_{i}).
\end{align}
On the other hand, under the constraints of the dispersion relations, it is easy to check that the following algebraic relation holds
\begin{align}
	\frac{q_{i}+q_{j}}{p_{i}-p_{j}}\frac{1}{4({p}_{i}{q}_{i}-{p}_{j}{q}_{j})}=\frac{p_{i}+p_{j}}{q_{i}-q_{j}}\frac{1}{4({p}_{i}{q}_{i}-{p}_{j}{q}_{j})}=\frac{1}{2\Big({({p}_{i}-{p}_{j})}^{2}+{({q}_{i}-{q}_{j})}^{2}\Big)}.
\end{align}
Thus by introducing auxiliary zero-valued Pfaffians, we have
\begin{equation}
	\begin{split}
		&\frac{1}{2}({d}_{2},{d}_{0},\bullet)({d}_{0},{d}_{0},\bullet)+\frac{1}{2}({d}_{0},{d}_{0},\bullet)({d}_{2},{d}_{0},\bullet)
		+\frac{1}{2}({d}_{-2},{d}_{0},\bullet)({d}_{0},{d}_{0},\bullet)+\frac{1}{2}({d}_{0},{d}_{0},\bullet)({d}_{-2},{d}_{0},\bullet)\\
		&-({d}_{1},{d}_{0},\bullet)({d}_{1},{d}_{0},\bullet)-({d}_{-1},{d}_{0},\bullet)({d}_{-1},{d}_{0},\bullet)\\
		&=\sum \limits_{j,l=1}^{2N}{(-1)}^{j+l}\bigg(\frac{1}{2}({d}_{2},{a}_{j})({d}_{0},{a}_{l})+\frac{1}{2}({d}_{0},{a}_{j})({d}_{2},{a}_{l})
		+\frac{1}{2}({d}_{-2},{a}_{j})({d}_{0},{a}_{l})+\frac{1}{2}({d}_{0},{a}_{j})({d}_{-2},{a}_{l})\\
		&\quad -({d}_{1},{a}_{j})({d}_{1},{a}_{l})-({d}_{-1},{a}_{j})({d}_{-1},{a}_{l})\bigg)({d}_{0},\hat{{a}_{j}})({d}_{0},\hat{{a}_{l}})\\
		&=\sum \limits_{j,l=1}^{2N}{(-1)}^{j+l}\bigg(\frac{{p}^{2}_{j}+{p}^{2}_{l}}{2}+\frac{{q}^{2}_{j}+{q}^{2}_{l}}{2}-{p}_{j}{p}_{l}-{q}_{j}{q}_{l}\bigg)e^{{\eta}_{j}+{\eta}_{l}}
		({d}_{0},\hat{{a}_{j}})({d}_{0},\hat{{a}_{l}})\\
		&=\sum \limits_{j,l=1}^{2N}{(-1)}^{j+l}\frac{2\Big({({p}_{j}-{p}_{l})}^{2}+{({q}_{j}-{q}_{l})}^{2}\Big)}{4}e^{{\eta}_{j}+{\eta}_{l}}
		({d}_{0},\hat{{a}_{j}})({d}_{0},\hat{{a}_{l}})\\
		&=\sum \limits_{j,l=1}^{2N}{(-1)}^{j+l}({p}_{j}{q}_{j}-{p}_{l}{q}_{l})({a}_{j},{a}_{l})({d}_{0},\hat{{a}_{j}})({d}_{0},\hat{{a}_{l}})\\
		&=\sum \limits_{l=1}^{2N}(-1)^{l}(-2{p}_{l}{q}_{l})({d}_{0},\hat{{a}_{l}})\sum \limits_{j=1}^{2N}(-1)^{j}({a}_{j},{a}_{l})({d}_{0},\hat{{a}_{j}})\\
		&=\sum \limits_{l=1}^{2N}(-1)^{l}(-2{p}_{l}{q}_{l})({d}_{0},\hat{{a}_{l}})\bigg(-({d}_{0},{a}_{l})(\bullet)+(-1)^{l+1}({a}_{l},{b}_{l})({d}_{0},\hat{{b}_{l}})\bigg)\\
		&=2\sum \limits_{l=1}^{2N}(-1)^{l}{p}_{l}{q}_{l}({d}_{0},{a}_{l})({d}_{0},\hat{{a}_{l}})(\bullet)+2\sum \limits_{l=1}^{2N}{p}_{l}{q}_{l}({a}_{l},{b}_{l})({d}_{0},\hat{{a}_{l}})({d}_{0},\hat{{b}_{l}})\\
		&=-2({d}_{0},\rho,\bullet)(\bullet)+\frac{I}{2}.
	\end{split}
\end{equation}
Then we deduce that
\begin{equation}
	4({d}_{0},\rho,\bullet)(\bullet)-2({d}_{1},{d}_{0},\bullet)^{2}-2({d}_{-1},{d}_{0},\bullet)^{2}=I,
\end{equation}
which gives the proof of the second bilinear identity and ends the proof.

\section{Multi-component coupled Higgs equation and its solution}

\subsection{one-soliton solution}
\quad \quad
In this section, we consider the explicit form of one soliton and two soliton solutions of the multi-component coupled Higgs equation with the aid of the Pfaffian expressions. The soliton dynamics are further analysed in this section.
\subsection{One-soliton solutions to the multi-component coupled Higgs equation}
We can give out the one soliton solutions to the multi-component coupled Higgs equation by expanding the following Pfaffians
\begin{align}
	f=(a_{1},a_{1}^{*},b_{1}^{*},b_{1})=1+t_{1\bar1}s_{1\bar1}e^{\eta_{1}+\eta_{\bar1}},\hspace{1em}g_{k}=(d_{0},\beta_{k},a_{1},a_{1}^{*},b_{1}^{*},b_{1})=\alpha_{1}^{(k)}e^{\eta_{1}}.
\end{align}
where we denote
\begin{align}\label{Eq-sij}
	s_{i\bar j}=\frac{r\sum_{k=1}^{M}\delta_{k}\alpha_{i}^{(k)}\bar\alpha_{j}^{(k)}}{4(p_{i}q_{i}-\bar p_{j}\bar q_{j})},\hspace{1em}t_{i\bar j}=\frac{p_{i}-\bar p_{j}}{q_{i}+\bar q_{j}}.
\end{align}
Thus the potentials read as
\begin{align}
	v=2(\ln f)_{xx}=2p_{1R}^2\sech^{2}\Big(\frac{\ln{(s_{1\bar 1}t_{1\bar 1})}+\eta_{1}+\bar\eta_{1}}{2}\Big),
\end{align}
and
\begin{align}
	&u_{k}=\frac{g_{k}}{f}=\frac{\alpha_{1}^{(k)}}{2\sqrt{t_{1\bar{1}}s_{1\bar{1}}}}\sech\Big(\frac{\ln{(s_{1\bar 1}t_{1\bar 1})}+\eta_{1}+\bar\eta_{1}}{2}\Big)e^{\mathrm{i}\eta_{1I}},\\
	&|u_{k}|^{2}=\frac{|\alpha_{1}^{(k)}|^{2}}{4t_{1\bar{1}}s_{1\bar{1}}}\sech^{2}\Big(\frac{\ln{(s_{1\bar 1}t_{1\bar 1})}+\eta_{1}+\bar\eta_{1}}{2}\Big),
\end{align}
where $\eta_{1I}$ refers to the imaginary part of $\eta_{1}$. We can know that the amplitude of $u_{k}$ and $v$ are $\frac{\sqrt{\frac{{(p_{1}+\bar{p}_{1})}^{2}+{(q_{1}+\bar{q}_{1})}^{2}}{2}}\alpha_{1}^{(k)}}{{\big(r\sum_{k=1}^{M}\delta_{k}|\alpha_{1}^{(k)}|^{2}\big)}^{\frac{1}{2}}}$and
$2p_{1R}^2$
respectively.
\subsection{Dynamic of two-soliton solutions to the two-component coupled Higgs equation}
We can obtain the expression of two-soliton solutions in the two components system through the Pfaffian expression of multi-component $N$-soliton solutions that
\begin{align}
	f
	&=1+\sum_{i,j=1}^{2}t_{i\bar j}s_{i\bar j}e^{\eta_{i}+\bar\eta_{j}}-(t_{12}t_{\bar1\bar2}-t_{1\bar1}t_{2\bar2}+t_{1\bar2}t_{2\bar1})(s_{1\bar1}s_{2\bar2}-s_{1\bar2}s_{2\bar1})e^{\eta_{1}+\eta_{2}+\bar\eta_{1}+\bar\eta_{2}},\\
	g_{1}
	&=\alpha_{1}^{(1)}e^{\eta_{1}}+\alpha_{2}^{(1)}e^{\eta_{2}}-(t_{12}-t_{1\bar2}+t_{2\bar2})(\alpha_{2}^{(1)}s_{1\bar2}-\alpha_{1}^{(1)}s_{2\bar2})e^{\eta_{1}+\eta_{2}+\bar\eta_{2}}\\\nonumber
	&\quad\quad-(t_{12}-t_{1\bar1}+t_{2\bar1})(\alpha_{2}^{(1)}s_{1\bar1}-\alpha_{1}^{(1)}s_{2\bar1})e^{\eta_{1}+\eta_{2}+\bar\eta_{1}},\\
	g_{2}
	&=\alpha_{1}^{(2)}e^{\eta_{1}}+\alpha_{2}^{(2)}e^{\eta_{2}}-(t_{12}-t_{1\bar2}+t_{2\bar2})(\alpha_{2}^{(2)}s_{1\bar2}-\alpha_{1}^{(2)}s_{2\bar2})e^{\eta_{1}+\eta_{2}+\bar\eta_{2}}\\\nonumber
	&\quad\quad-(t_{12}-t_{1\bar1}+t_{2\bar1})(\alpha_{2}^{(2)}s_{1\bar1}-\alpha_{1}^{(2)}s_{2\bar1})e^{\eta_{1}+\eta_{2}+\bar\eta_{1}}.
\end{align}
That is 
\begin{equation}
	\begin{split}
		g_{k}=\sum_{i=1}^{2}\alpha_{i}^{(k)}e^{\eta_{i}}-\sum_{j=1}^{2}(t_{12}-t_{1\bar{j}}+t_{2\bar{j}})(\alpha_{2}^{(k)}s_{1\bar{j}}-\alpha_{1}^{(k)}s_{2\bar{j}})e^{\eta_{1}+\eta_{2}+\bar{\eta}_j},\hspace{1.285714em}k=1,2
	\end{split}
\end{equation}
in general, where $s_{ij}$ and $t_{ij}$ are defined in (\ref{Eq-sij}).

We study the asymptotic behaviours to explain interaction properties. We assume that the real parts of $p_i$ and $q_i$ are all positive, $|q_{1}|\neq|q_{2}|$ and $|p_{i}|,|q_{i}|>0$. Then we have the following two situations:
\begin{equation}
	(i)\quad\quad  \eta_{1R}=O(1), \quad \eta_{2R}\rightarrow\pm\infty , \quad  t\rightarrow\pm\infty;
\end{equation}
\begin{equation}
	(ii)\quad\quad  \eta_{2R}=O(1), \quad \eta_{1R}\rightarrow\pm\infty , \quad  t\rightarrow\pm\infty.
\end{equation}
Then we have the following asymptotic form for $v$ and $u_{k}$
$$
v\sim\left\{
\begin{aligned}
	&2p^{2}_{1R}\sech^{2}\Big(\eta_{1R}+\frac{r_{1\bar{1}}}{2}\Big),\hspace{5em}\eta_{2R}\rightarrow -\infty,\eta_{1R}\sim O(1)\\
	&2p^{2}_{1R}\sech^{2}\Big(\eta_{1R}+\frac{r_{12\bar{1}\bar{2}}-r_{2\bar{2}}}{2}\Big),\hspace{2em}\eta_{2R}\rightarrow +\infty,\eta_{1R}\sim O(1) \\
	&2p^{2}_{2R}\sech^{2}\Big(\eta_{2R}+\frac{r_{2\bar{2}}}{2}\Big),\hspace{5em}\eta_{1R}\rightarrow -\infty,\eta_{2R}\sim O(1)\\
	&2p^{2}_{2R}\sech^{2}\Big(\eta_{2R}+\frac{r_{12\bar{1}\bar{2}}-r_{1\bar{1}}}{2}\Big),\hspace{2em}\eta_{1R}\rightarrow +\infty,\eta_{2R}\sim O(1)
\end{aligned}
\right.
$$

\begin{figure}
	\begin{center}
		\begin{tabular}{ccc}
			\includegraphics[height=0.380\textwidth,angle=0]{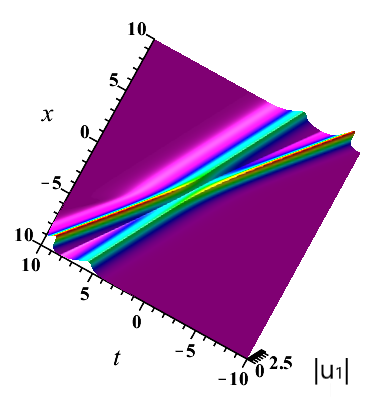} &
			\includegraphics[height=0.380\textwidth,angle=0]{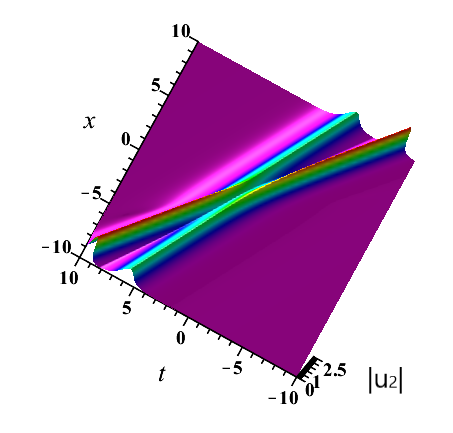} &
		\end{tabular}
	\end{center}
	\caption{Elastic two-soliton interaction in the two-component coupled Higgs equation when $ p_{1}=1+\mathrm{i},  p_{2}=2.5+1.7\mathrm{i}, \alpha_{1}^{(1)}=1, \alpha_{1}^{(2)}=1, \alpha_{2}^{(1)}=0.9, \alpha_{2}^{(2)}=0.9, c=3, r=2, \delta_{1}=\delta_{2}=1$}
	\label{Fig-multi-ration-plots5}
\end{figure}
\begin{figure}
	\begin{center}
		\begin{tabular}{ccc}
			\includegraphics[height=0.380\textwidth,angle=0]{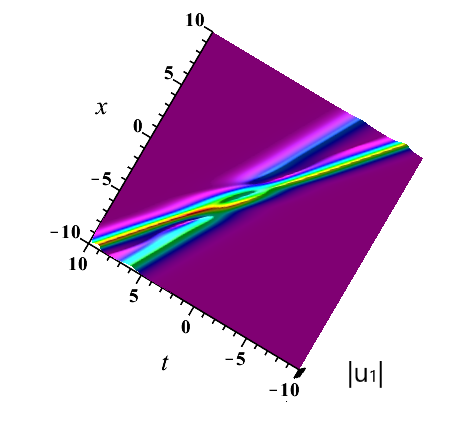} &
			\includegraphics[height=0.380\textwidth,angle=0]{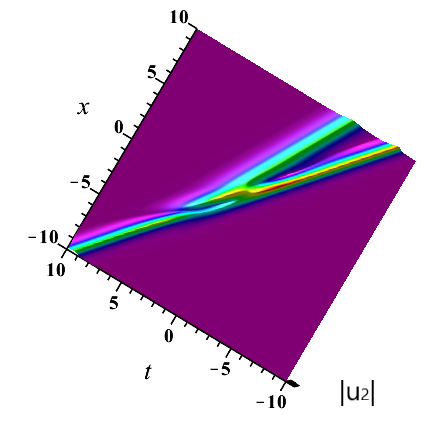} &
		\end{tabular}
	\end{center}
	\caption{Y-shaped inelastic collision in the two-component coupled Higgs equation when $ p_{1}=1+\mathrm{i},  p_{2}=2.5+1.7\mathrm{i}, \alpha_{1}^{(1)}=1, \alpha_{1}^{(2)}=1, \alpha_{2}^{(1)}=0, \alpha_{2}^{(2)}=0.9, c=3, r=2, \delta_{1}=\delta_{2}=1$}
	\label{Fig-multi-ration-plots6}
\end{figure}
\begin{figure}
	\begin{center}
		\begin{tabular}{ccc}
			\includegraphics[height=0.380\textwidth,angle=0]{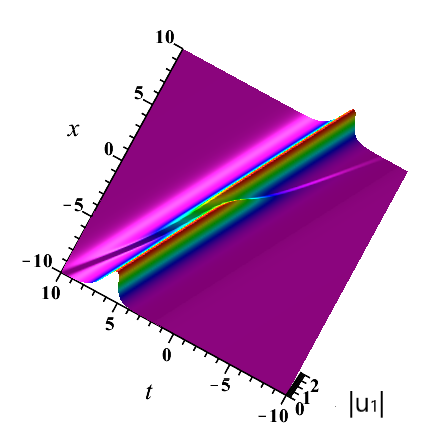} &
			\includegraphics[height=0.380\textwidth,angle=0]{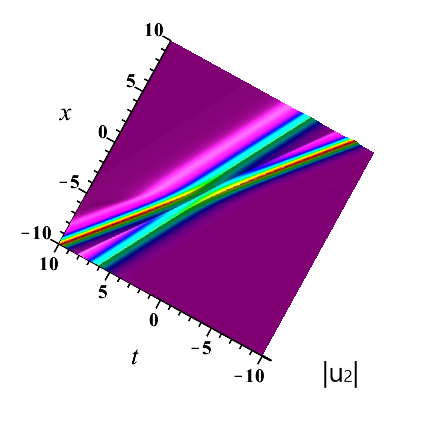} &
		\end{tabular}
	\end{center}
	\caption{Collision of two solitons to the two-component coupled Higgs equation when $ p_{1}=1+\mathrm{i},  p_{2}=2.5+1.7\mathrm{i}, \alpha_{1}^{(1)}=1, \alpha_{1}^{(2)}=0, \alpha_{2}^{(1)}=0, \alpha_{2}^{(2)}=0.9, c=3, r=2, \delta_{1}=\delta_{2}=1$}
	\label{Fig-multi-ration-plots7}
\end{figure}
and
$$ u_{k}\sim\left\{
\begin{aligned}
	&\frac{a^{(k)}_{1}}{2\sqrt{a_{1\bar{1}}}}\sech\Big(\eta_{1R}+\frac{r_{1\bar{1}}}{2}\Big)e^{\mathrm{i}\eta_{1I}},\hspace{10em}\eta_{2R}\rightarrow -\infty,\eta_{1R}\sim O(1)\\
	&\frac{a^{(k)}_{12\bar{2}}}{2\sqrt{a_{2\bar{2}}a_{12\bar{1}\bar{2}}}}\sech\Big(\eta_{1R}+\frac{r_{12\bar{1}\bar{2}}-r_{2\bar{2}}}{2}\Big)e^{\mathrm{i}\eta_{1I}},\hspace{5.6em}\eta_{2R}\rightarrow +\infty,\eta_{1R}\sim O(1) \\
	&\frac{a^{(k)}_{2}}{2\sqrt{a_{2\bar{2}}}}\sech\Big(\eta_{2R}+\frac{r_{2\bar{2}}}{2}\Big)e^{\mathrm{i}\eta_{2I}},\hspace{10em}\eta_{1R}\rightarrow -\infty,\eta_{2R}\sim O(1)\\
	&\frac{a^{(k)}_{12\bar{1}}}{2\sqrt{a_{1\bar{1}}a_{12\bar{1}\bar{2}}}}\sech\Big(\eta_{2R}+\frac{r_{12\bar{1}\bar{2}}-r_{1\bar{1}}}{2}\Big)e^{\mathrm{i}\eta_{2I}},\hspace{5.6em}\eta_{1R}\rightarrow +\infty,\eta_{2R}\sim O(1)
\end{aligned}
\right.
$$
where we denote
\begin{align}
	&e^{r_{i\bar{j}}}=a_{i\bar{j}}=t_{i\bar{j}}s_{i\bar{j}},\\
	&a_{12\bar{j}}^{(k)}=-(t_{12}-t_{1\bar{j}}+t_{2\bar{j}})(\alpha^{(k)}_{2}s_{1\bar{j}}-\alpha^{(k)}_{1}s_{2\bar{j}}),\\
	&e^{r_{12\bar{1}\bar{2}}}=a_{12\bar{1}\bar{2}}=-(t_{12}t_{\bar{1}\bar{2}}-t_{1\bar{1}}t_{2\bar{2}}+t_{1\bar{2}}t_{2\bar{1}})(s_{1\bar{1}}s_{2\bar{2}}-s_{1\bar{2}}s_{2\bar{1}}),
\end{align}
for simplicity. We can rule out that the velocity and amplitude of $v$ stay the same before and after the collision from the asymptotic behaviours. The collisions for $v$ are absolutely elastic. The phase shifts due to interaction for $v$ and $u_{k}$ are $\frac{r_{12\bar{1}\bar{2}}-r_{1\bar{1}}-r_{2\bar{2}}}{2}$. However the collisions for $u_{k}$ are not always elastic.
When $\frac{\alpha_{1}^{(1)}}{\alpha_{1}^{(2)}}=\frac{\alpha_{2}^{(1)}}{\alpha_{2}^{(2)}}$, two components of solitons collide without energy exchange. As is shown in Fig. \ref{Fig-multi-ration-plots5}, setting the parameters $  p_{1}=1+\mathrm{i},  p_{2}=2.5+1.7\mathrm{i}, \alpha_{1}^{(1)}=1, \alpha_{1}^{(2)}=1, \alpha_{2}^{(1)}=0.9, \alpha_{2}^{(2)}=0.9, c=3, r=2, \delta_{1}=\delta_{2}=1$, we see the elastic collision of the two solitons. The two different solitons collide and then travel apart from each other. No velocities and amplitudes change in this case and hence the collision is elastic.
Similar as some other coupled integrable systems \cite{6,7}, the interaction may cause the change of the amplitude of each soliton if the parameters $\alpha_{i}^{(k)}(k=1,2)$ take different values.
If the parameters $\alpha_{i}^{(k)}(k=1,2)$ are not in proportion, in most cases the energy exchange might exists during the collision. As is depicted in Fig. \ref{Fig-multi-ration-plots6}, setting parameters as $  p_{1}=1+\mathrm{i}, p_{2}=2.5+1.7\mathrm{i}, \alpha_{1}^{(1)}=1, \alpha_{1}^{(2)}=1, \alpha_{2}^{(1)}=0, \alpha_{2}^{(2)}=0.9, c=3, r=2,\delta_{1}=\delta_{2}=1$, we find that one soliton merges into the other in the profile of $|u_{2}|$. Though there is an energy change, there still remains two solitons in $|u_{1}|$, which means we find that the energy is packed in only one soliton while the energy of the other soliton is not zero.
When we take the parameters as $ p_{1}=1+\mathrm{i}, p_{2}=2.5+1.7\mathrm{i}, \alpha_{1}^{(1)}=1, \alpha_{1}^{(2)}=0, \alpha_{2}^{(1)}=0, \alpha_{2}^{(2)}=0.9, c=3, r=2, \delta_{1}=\delta_{2}=1$, we see only one soliton with a phase shift in $|u_{1}|$ as is shown in Fig. \ref{Fig-multi-ration-plots7}. The two solitons exist in the second component $|u_2|$ and the collision is elastic with only a phase shift.
\section{Conclusion}
In this work, we consider the coupled Higgs equation via Hirota's bilinear method. We give the multi-component generalization to the coupled Higgs equation. One and two soliton solutions of the single-component system are obtained directly via Hirota's method and $N$-soliton solutions of the multi-component system are found by Pfaffian techniques. The dynamic behaviours of both the single-component and multi-component systems are investigated. Parallel solitons, two solitons traveling in the opposite directions, elastic and inelastic collisions, periodic and nearly periodic collisions are found in the solutions of the coupled Higgs equation.

We notice the Lax pair of the coupled Higgs equation is not covered in literatures. Meanwhile, rogue wave dynamics in the coupled Higgs equation is still unknown yet. It is nature for us to consider the modulation instabilities and the rogue waves in the coupled Higgs equation. Furthermore, since that the coupled Higgs equation has doubly periodic solutions in the forms of elliptic functions, it is also worth considering the rogue waves on the elliptic backgrounds for the coupled Higgs equation. These will be considered in our further work.

\section*{\bf Acknowledgements}
I would like to appreciate Professor Guo-Fu Yu for his discussion in this work.


\vskip .5cm

\begin{thebibliography}{00}
	\bibitem{BE1} Y.V. Bludov, V.V. Konotop and N. Akhmediev, Matter rogue waves. {\it Phys. Rev. A.} 80 (2009) 033610.
	
	\bibitem{BE2} Y.V. Bludov, V.V. Konotop and N. Akhmediev, Rogue waves as patial energy concentrators in arrays of nonlinear waveguides. {\it Opt. Lett.} 34 (2009) 3015-3017.
	
	\bibitem{media} T.P. Horikis and M.J. Ablowitz, Rogue waves in nonlocal media. {\it Phys. Rev. E} 95 (2017) 042211.
	
	\bibitem{superfluid} A.N. Ganshin, V.B. Efimov, G.V. Kolmakov, L.P. Mezhov-Deglin and P.V.E. McClintock, Observation of an Inverse Energy Cascade in Developed Acoustic Turbulence in Superfluid Helium. {\it Phys. Rev. Lett.} 101 (2008) 065303.
	
	\bibitem{plasma1} W.M. Moslem, Erratum: "Langmuir rogue waves in electronpositron plasmas". {\it Phys. Plasmas} 18 (2011) 032301.
	
	\bibitem{plasma2} H. Bailung, S.K. Sharma and Y. Nakamura, Observation of Peregrine Solitons in a Multicomponent Plasma with Negative Ions. {\it Phys. Rev. Lett.} 107 (2011) 255005.
	
	\bibitem{optical1} D.R. Solli, C. Ropers, P. Koonath and B. Jalali, Optical rogue waves. {\it Nature} 450 (2007) 1054-1057.
	
	\bibitem{optical2} B. Kibler, J. Fatome, C. Finot, G. Millot, F. Dias, G. Genty, N. Akhmediev and J.M. Dudley, The Peregrine soliton in nonlinear fibre optics. {\it Nature Phys.} 6 (2010) 790-795.
	
	\bibitem{DT} C.H. Gu, H.S. Hu and Z.X. Zhou, Darboux transformations in integrable systems: theory and their applications to geometry. {\it Springer,} (2006).
	
	\bibitem{DT1} H.Q. Zhang and F. Chen, Dark and antidark solitons for the defocusing coupled Sasa-Satsuma system by the Darboux transformation. {\it Appl. Math. Lett.} 88 (2019) 237-242.
	
	\bibitem{BM} R. Hirota, The Direct Method in Soliton Theory. {\it Cambridge University Press, Cambridge,} (2004).
	
	\bibitem{BM1} X.B. Hu, C.X. Li, J.J.C. Nimmo and G.F. Yu, An integrable symmetric (2+1)-dimensional Lotka-Volterra equation and a family of its solutions. {\it J. Phys. A} 38 (2005) 195-204.
	
	\bibitem{IS} M.J. Ablowitz and P.A. Clarkson, Solitons, Nonlinear Evolution Equations and Inverse Scattering. {\it Cambridge University Press, Cambridge,} (1991).
	
	\bibitem{RH} M.J. Ablowitz and A.S. Fokas, Complex Variables: Introduction and Applications. {\it Cambridge University Press, Cambridge,} (1997).
	
	\bibitem{Tajiri1} M. Tajiri, On $N$-soliton solutions of coupled Higgs field equation. {\it J. Phys. Soc. Jpn.} 52 (1983) 2277-2280.
	
	\bibitem{V} V.V. Tsegel'nik, Self-similar solutions of a system of two nonlinear partial differential equations. {\it Diff. Equ.} 36 (2000) 480-482.
	
	\bibitem{Tajiri2} M. Tajiri, On soliton solutions of the nonlinear coupled Klein-Gordon equation. {\it J. Phys. Soc. Jpn.} 52 (1983) 3722-3726.
	
	\bibitem{Tajiri3} M. Tajiri and M. Hagiwara, Similarity solutions of the two-dimensional coupled nonlinear Schr\"odinger equation. {\it J. Phys. Soc. Jpn.} 52 (1983) 3727-3734.
	
	\bibitem{DS} A. Davey and K. Stewartson, On three-dimensional packets of surface waves. {\it Proc. Roy. Soc. London Ser. A} 338 (1974) 101-110.
	
	\bibitem{DS1} X.Z. Hao, Y.P. Liu, X.Y. Tang and Z.B. Li, The residual symmetry and exact solutions of the Davey-Stewartson III equation. {\it Comput. Math. Appl.} 73 (2017) 2404-2414.
	
	\bibitem{DS2} Y.B. Liu, A.S. Fokas, D. Mihalache and J.S. He, Parallel line rogue waves of the third-type Davey-Stewartson equation. {\it Rom. Rep. Phys.} 68 (2016) 1425-1446.
	
	\bibitem{DS3} J.G. Rao, K. Porsezian and J.S. He, Semi-rational solutions of the third-type Davey-Stewartson equation. {\it Chaos} 27 (2017) 083115.
	
	\bibitem{Hu} X.B. Hu, B.L. Guo and H.W. Tam, Homoclinic orbits for the coupled Schr\"odinger-Boussinesq equation and coupled Higgs equation. {\it J. Phys. Soc. Jpn.} 72 (2003) 189-190.
	
	\bibitem{exp-function} H. Manafian, Jalil, Zamanpour and Isa, Analytical treatment of the coupled Higgs equation and the Maccari system via exp-function method. {\it Acta Univ. Apulensis Math. Inform.} 33 (2013) 203-216.
	
	\bibitem{generalizedalgebraicmethod} H. Zhao, Applications of the generalized algebraic method to special-type nonlinear equations. {\it Chao Solitons Fractals} 36 (2008) 359-369.
	
	\bibitem{exactsolutions} Y.C. Hon and E.G. Fan, A series of exact solutions for coupled Higgs field equation and coupled Schr\"{o}dinger-Boussinesq equation. {\it Nonlinear Anal.} 71 (2009) 3501-3508.
	
	\bibitem{G1} A. Jabbari, H. Kheiri and A. Bekir, Exact solutions of the coupled Higgs equation and the Maccari system using He's semi-inverse method and $(G'/G)$-expansion method. {\it Comput. Math. Appl.} 62 (2011) 2177-2186.
	
	\bibitem{G2} M.N. Alam, M.G. Hafez, F.B.M. Belgacem and M.A. Akbar, Applications of the novel $(G'/G)$ expansion method to find new exact traveling wave solutions of the nonlinear coupled Higgs field equations. {\it Nonlinear Stud.} 22 (2015) 613-633.
	
	\bibitem{standingwave1} E.G. Fan, K.W. Chow and J.H. Li, On doubly periodic standing wave solutions of the coupled Higgs field equation. {\it Stud. Appl. Math.} 128 (2012) 86-105.
	
	\bibitem{standingwave2} G.Q. Xu, New types of doubly periodic standing wave solutions for the coupled Higgs field equation. {\it Abstr. Appl. Anal.} (2014) 769561.
	
	\bibitem{Bifurcationsoftravelingwave} S.Q. Tang and S. Xia, Bifurcations of traveling wave solutions for the coupled Higgs field equation. {\it Int. J. Differ. Equ.} (2011) 547617.
	
	\bibitem{Backlundtransformation} A.H. Arnous, M. Mirzazadeh and M. Eslami, The B\"{a}cklund transformation method of Riccati equation applied to coupled Higgs field and Hamiltonian amplitude equations. {\it Comput. Methods Differ. Equ.} 2 (2014) 216-226.
	
	\bibitem{1} R. Hirota, Soliton solutions to the BKP equations. I. The Pfaffian technique. {\it J. Phys. Soc. Japan} 58 (1989) 2285-2296.
	
	\bibitem{2} R. Hirota, Soliton solutions to the BKP equations. II. The integral equation. {\it J. Phys. Soc. Japan} 58 (1989) 2705-2712.
	
	\bibitem{3} M. Iwao and R. Hirota, Soliton solutions of a coupled modified KdV equations. {\it J. Phys. Soc. Japan} 66 (1997) 577-588.
	
	\bibitem{4} R. Hirota, M. Iwao and S. Tsujimoto, Soliton equations exhibiting Pfaffian solutions. Integrable systems: linear and nonlinear dynamics. {\it Glasg. Math. J.} 43A (2001) 33-41.
	
	\bibitem{5} Y. Ohta, Discretization of coupled nonlinear Schr\"odinger equations. {\it Stud. Appl. Math.} 122 (2009) 427-447.
	
	\bibitem{6} Z.W. Xu, G.F. Yu and Z.N. Zhu, Soliton dynamics to the multi-component complex coupled integrable dispersionless equation. {\it Commun. Nonlinear Sci. Numer. Simul.} 40 (2016) 28-43.
	
	\bibitem{7} G.F. Yu and Z.W. Xu, Dynamics of a differential-difference integrable (2+1)-dimensional system. {\it Phys. Rev. E(3)} 91 (2015) 062902.
	
	
\end{thebibliography}
\end{document}